\documentclass{article}

\usepackage{arxiv}
\usepackage[utf8]{inputenc} 
\usepackage[T1]{fontenc}    
\usepackage{hyperref}       
\usepackage{url}            
\usepackage{booktabs}       
\usepackage{amsfonts}       
\usepackage{nicefrac}       
\usepackage{microtype}      
\usepackage{lipsum}		
\usepackage{graphicx}
\usepackage[numbers]{natbib}
\usepackage{doi}
\usepackage{siunitx}
\usepackage{tikz}
\usetikzlibrary{arrows}
\usepackage{amsmath}
\usepackage[labelfont=bf,textfont=bf]{caption}

\usepackage{accents}
\newcommand{\ubar}[1]{\underaccent{\bar}{#1}}

\newcommand{\vect}[1]{\mathbf{#1}}
\DeclareMathOperator{\rank}{rank}
\newcommand{\determ}[1]{\mathrm{det}(#1)} 
\DeclareMathOperator{\diag}{diag}
\DeclareMathOperator{\sech}{sech}

\usepackage[standard]{ntheorem}

\newcommand{\qed}{\tag*{$\blacksquare$}}
\def\doubleunderline#1{\underline{\underline{#1}}}

\title{Neural-inspired Measurement Observability}

\date{June 5, 2022}

\author{ {Burak Boyac\i o\u{g}lu, Alice C. Schwarze, Bingni W. Brunton, and Kristi A. Morgansen\thanks{Burak Boyac\i o\u{g}lu and Kristi A. Morgansen are with the William E. Boeing Department of Aeronautics and Astronautics, University of Washington, Seattle, WA 98195-2400, USA. Alice C. Schwarze and Bingni W. Brunton are with the Department of Biology, University of Washington, Seattle, WA 98195-1800, USA. \texttt{burakb@uw.edu}, \texttt{morgansn@uw.edu} \texttt{alice.c.schwarze@dartmouth.edu}, \texttt{bbrunton@uw.edu}}} \\
University of Washington, Seattle, WA 98195
}

\hypersetup{
pdftitle={Neural-inspired Measurement Observability},
pdfauthor={Burak Boyac\i o\u{g}lu},
}

\begin{document}
\maketitle

\begin{abstract}
The neural encoding by biological sensors of flying insects, which prefilters stimulus data before sending it to the central nervous system in the form of voltage spikes, enables sensing capabilities that are computationally low-cost while also being highly robust to noise. This process, which can be modeled as the composition of a linear moving average filter and a nonlinear decision function, inspired the work reported here to improve engineered sensing performance by maximizing the observability of particular neural-inspired composite measurement functions. We first present a tool to determine the observability of a linear system with measurement delay (the first element of the composition), then use a Lie algebraic observability approach to study nonlinear autonomous systems with output delay (the second element of the composition). The Lie algebraic tools are then extended to address overall observability of systems with composite outputs as in the neural encoder model we adopt. The analytical outcomes are supported using the empirical observability Gramian, and optimal sensor placement on a bioinspired wing model is performed using metrics based on the empirical Gramian.
\end{abstract}

\keywords{Neural encoding \and Nonlinear observability \and Sensor Placement}

\section{Introduction}

Flying insects  have the same needs for position and orientation awareness relative to the surrounding environment (pose) as engineered flight vehicles, but they must determine this information with much less computational power and much simpler sensors than the gyroscopes, GPS, accelerometers, and other modern sensors used in air and space vehicles.  By studying underlying principles of sensing mechanics and signal processing in biological flight, the potential exists to translate these basic principles to significantly more effective engineered systems.  Here, we consider the use of mechanosensing in insects via strain measurements and neural processing of the resulting strain signals to inform engineered sensing capabilities.

Sensory neurons in an animal's body are responsible for receiving stimulus information and delivering it to the rest of the nervous system \cite{dayan2001theoretical}. This data transfer relies on action potentials (voltage spikes) \cite{dayan2001theoretical}. Neural encoding is the mechanism that converts a stimulus into these voltage spikes. This mapping is not fully understood and might be so complex that, for example, a single unit in an animal's retina might be specialized in responding to particular light patterns and to being insensitive to others \cite{GOLLISCH2010}.
Here, the sensing modality on which we will focus is that of mechanosensing. Campaniform sensilla are a type of strain-sensitive mechanoreceptors found in flying insects and use a signal history no longer than $\SI{40}{\milli\second}$ \cite{Fox2010}. Inertial rotations are detectable from this strain information \cite{dickerson2014, eberle2015} at a rate much faster than the essential but relatively slow visual system \cite{pratt2017} thus enabling rapid maneuvers of insects that require sensory feedback. In \cite{mohren2018}, it is shown that small numbers of sensors inspired by the campaniform sensilla placed in advantageous locations can provide a good classification of strain data from a hawkmoth \textit{Manduca sexta} wing model for the purpose of determining animal rotation rate. Spatial distribution of campaniform sensilla on insect wings in terms of form and function has been further discussed in \cite{AIELLO20218}.

A simple model of neural encoding can be obtained by composing a linear filter and a nonlinear decision function. This temporal filter acts like a moving average filter in engineered systems, and the output at any given time is not just a function of the state at that time but is also an explicit function of previous states, that is, neural encoded measurements use the history of the stimulus. The range of filter length is nominally unlimited, however, effective stimulus history for spiking is typically a few hundred milliseconds long in practice \cite{dayan2001theoretical}.

To explore these ideas from an engineering perspective, we here consider neural processing and mechanosensing using systems theoretic tools.  In engineering systems, the ability to reconstruct state data from measurements is assessed by the system observability, and the algorithms and devices used to perform the reconstruction are implemented via filters. Since the original introduction of filter design for linear systems such as the Luenberger filter and the Kalman filter \cite{KALMAN1960}, the assessment of observability has typically been handled independently from filter design with the idea that any improvements in observability will necessarily benefit any filtering methods used.   In nonlinear systems, observability analysis is generally approached using differential geometric methods, and, unlike linear systems where the separation principle guarantees that system actuation has no effect on sensing ability, system actuation in nonlinear systems can affect whether and how well system sensing can be leveraged for state reconstruction \cite{Hermann1977}. For systems where analytical tools are impractical, the empirical observability Gramian, a numerical tool first introduced by Moore \cite{mooore1981} then systematized in \cite{lall1999} for model reduction, can be adopted for local observability analysis. Recently, an empirical Gramian rank condition for weak observability of nonlinear systems and an equivalent sufficient condition based on the minimum singular value of the empirical observability Gramian were developed in \cite{Powel2015}. A numerical approximation to obtain a lower bound for the minimum singular value was presented in \cite{powelArXiv}, as the analytical calculations are usually intractable. Finally, a powerful open-source toolbox for the calculation of empirical Gramians, \textit{emgr}, has been introduced for both time-varying and time-invariant systems \cite{himpe2018}.

The observability of four common neuron models without output delay and with one measured state was studied in \cite{Aguirre2017}. Observability of linear time-invariant systems with output delay was discussed and observer design was given for a class of such systems in \cite{Olbrot1981}. Recently, a new notion of observability was introduced for systems with commensurate delays in dynamics and measurements \cite{Califano2020, Califano2021}.

Here, we assume that only the neural encoded data is available to process, not the raw strain information, to be consistent with typical data obtained from biological systems. Existing analytical tools are revisited and developed for systems with output delay. To study the observability of systems with composite output functions, we propose an expression of higher Lie derivatives of a function composition with respect to a vector field. We then use the empirical Gramian for the observability analysis of an Euler-Lagrange model of hawkmoth \textit{Manduca sexta} flapping wing dynamics with a neural-inspired output function which has a delay and also can be written as a composition of two functions. Finally, a linear combination of two Gramian-based unobservability measures is used to determine optimal sensor placement throughout the veins on the wing, and the effect of neural encoding parameters on observability is explored.

The remainder of the paper has been organized as follows.
Section \ref{sec:model} describes the wing model with the adopted neural encoding model. Section~\ref{sec:back} summarizes existing observability analysis tools.
Section~\ref{sec:delay} and \ref{sec:compost} introduce the developed tools for observability analysis of systems with output delay and systems with a composite output function, respectively.
Empirical Gramian-based simulation results are included in Section~\ref{sec:sims}.
Finally, Section~\ref{sec:last} gives the conclusions and directions for future work.

\section{System Model}\label{sec:model}

 The focus of the work in this paper is the study of sensing based on neural encoded strain measurements in flapping wing flight.  While the lifting surfaces of the system are assumed to undergo cyclic flapping, the results translate to any lifting surfaces subject to structural loading. The neural encoding model has two key features that we intend to explore in this paper: delay in measurements and composition of functions in measurements.
 
\subsection{Wing Model Dynamics}

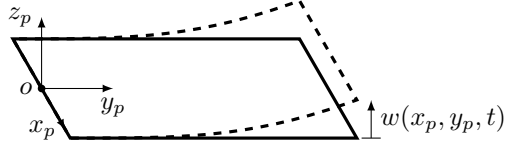
\begin{figure}[hbt!]
\begin{center}
\begin{tikzpicture}[>=latex]
 \draw[->] (0,0) -- +(0.375in,0) node[anchor=north]{$y_p$};
 \draw[->] (0,0) -- +(0,0.375in) node[anchor=east]{$z_p$};
 \draw[->] (0,0) -- +(-60:0.25in) node[anchor=east]{$x_p$};
 \fill (0,0) circle(0.02in) node[anchor=east]{$o$}; 
 \draw[very thick] (0,0) -- ++(120:0.3in) node (n1){} -- ++(1.5in,0) node (n4){} -- ++(-60:0.6in) node (n3){} -- ++(-1.5in,0) node (n2){} -- cycle;
 \draw (n4.center) ++(0,0.2in) node (d4){};
 \draw (n3.center) ++(0,0.2in) node (d3){};
 \draw[very thick,dashed] (0,0) -- (n1.center) to[out=0,in=200] (d4.center) -- (d3.center) to[out=200,in=0] (n2.center) -- cycle;
 \draw (n3.center) ++(0.025in,0) -- ++(0.1in,0);
 \draw[->] (n3.center) ++(0.075in,0) -- ++(0,0.2in) node[midway,right]{$w(x_p,y_p,t)$};
\end{tikzpicture}
\caption{Free body diagram of a thin plate with out-of-plane bending in a rotating, accelerating reference frame, simplified from \cite{hinson2015}.}\label{fig:body}
\end{center}
\end{figure}

The particular system on which we base our work is the hawkmoth {\em Manduca sexta}. First, the dynamic model of a flexible, thin plate in a rotating, accelerating reference frame illustrated in Fig. \ref{fig:body} is obtained. Then the planform geometry of the hawkmoth wing is used for simulations. Here, we will summarize the results from the Euler-Lagrange modeling in \cite{hinson2015}. Details of the dynamic model can be found in the original paper.

\begin{figure}[hbt!]
\begin{center}
\begin{tikzpicture}[>=latex]
\matrix[column sep=0.2in]
{
\draw[->] (0,0) -- ++(0.5in,0) node[anchor=west]{$y_w$};
\draw[->] (0,0) -- ++(0,-0.5in) node[anchor=east]{$x_w$};
\fill (0,0) circle(0.03in) node[anchor=east]{$o$};
\draw[thick,scale=0.75] (0,0) to[out=30,in=180,looseness=0.5] (35mm,4mm) to[out=0,in=90,looseness=0.65] (50mm,-3mm) to[out=-90,in=0,looseness=0.5] (27mm,-16mm) to[out=180,in=-30,looseness=0.5] (5mm,-8mm) to[out=150,in=-90] (0,0);
\begin{scope}[rotate=20]
\draw[->] (0,0) -- ++(0.5in,0) node[anchor=west]{$y_w^\prime$};
\draw[->] (0,0) -- ++(0,-0.5in) node[anchor=west]{$x_w^\prime$};
\draw[thick,dashed,scale=0.75] (0,0) to[out=30,in=180,looseness=0.5] (35mm,4mm) to[out=0,in=90,looseness=0.65] (50mm,-3mm) to[out=-90,in=0,looseness=0.5] (27mm,-16mm) to[out=180,in=-30,looseness=0.5] (5mm,-8mm) to[out=150,in=-90] (0,0);
\end{scope}
\draw[->] (-90:0.4in) arc(-90:-70:0.4in);
\draw (-80:0.4in) node[anchor=north]{$\psi$};&

\draw[->] (0,0) -- ++(0.5in,0) node[anchor=north]{$y_w^\prime$};
\draw[->] (0,0) -- ++(0,0.5in) node[anchor=west]{$z_w^\prime$}; 
\fill (0,0) circle(0.03in) node[anchor=east]{$o$}; 
\draw[thick] (0,0) -- ++(37mm,0) node[near end,below]{\scriptsize{wing stroke plane}};
\begin{scope}[rotate=20]
\draw[->] (0,0) -- ++(0.5in,0) node[anchor=south]{$y_w^{\prime\prime}$};
\draw[->] (0,0) -- ++(0,0.5in) node[anchor=east]{$z_w^{\prime\prime}$}; 
\draw[thick,dashed] (0,0) -- ++(37mm,0);
\end{scope}
\draw[->] (0:0.4in) arc(0:20:0.4in);
\draw (10:0.4in) node[anchor=west]{$\theta$};&

\draw[->] (0,0) -- ++(0.5in,0) node[anchor=south]{$x_w^{\prime\prime}$};
\draw[->] (0,0) -- ++(0,0.5in) node[anchor=east]{$z_w^{\prime\prime}$};
\fill (0,0) circle(0.03in) node[anchor=north]{$o$};
\draw[thick] (-3mm,0) -- (12mm,0);
\begin{scope}[rotate=-20]
\draw[->] (0,0) -- ++(0.5in,0) node[anchor=north west]{$x_p$};
\draw[->] (0,0) -- ++(0,0.5in) node[anchor=west]{$z_p$};
\draw[thick,dashed] (-3mm,0) -- (12mm,0)  node[near end,below,rotate=-20]{\scriptsize{chord}};
\end{scope}
\draw[->] (0:0.4in) arc(0:-20:0.4in);
\draw (-10:0.4in) node[anchor=west]{$\alpha$};\\

\draw (18.75mm,0) node{position angle};&
\draw (18mm,0) node{elevation angle};&
\draw (0.25in,0) node{feathering angle};\\
};
\end{tikzpicture}
\caption{The Euler angles, $\psi$, $\theta$, and $\alpha$ that represent the wing stroke kinematics within the wing frame, $[x_w,y_w,z_w]^\top$ \cite{hinson2015}.}
\label{eulerAng}
\end{center}
\end{figure}
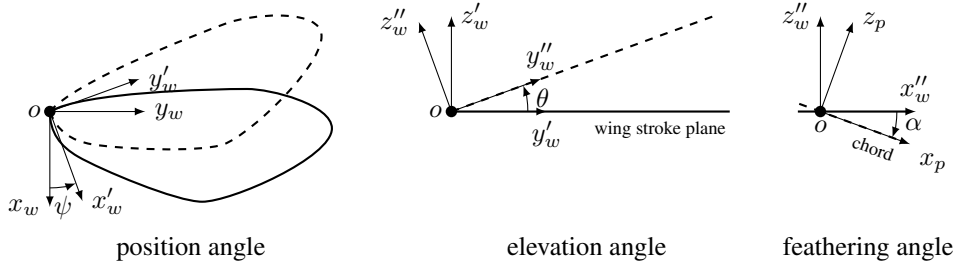

The out-of-plane deformation as a function of time and spatial coordinates on the  wing is described by
\begin{equation}
 w(x_p,y_p,t) = \sum_{i=1}^{n_m} \phi_i(x_p,y_p)\eta_i(t),
\label{eqn:weta}
\end{equation}
where $n_m$ is the chosen number of modes, and $w \in \mathbb{R}$. The terms $\phi_i(x_p,y_p) \in \mathbb{R}$ and $\eta_i(t) \in \mathbb{R}$ are the free-vibration mode shapes and modal coordinates, respectively.  The modes can either be derived analytically such as for the cantilever structure shown in the image or produced numerically through finite element analysis methods. The position, elevation and feathering Euler angles, illustrated in Fig. \ref{eulerAng}, are denoted by $\psi$, $\theta$ and $\alpha$, respectively. Finally, the axis of feathering rotation relative to the wing coordinate system origin is given as $\begin{bmatrix}0&0&x_r\end{bmatrix}^\top$.

In the dynamics model here, aerodynamic terms are neglected, which is a reasonable assumption since the aerodynamics are not a significant factor of the motion of a hawkmoth as shown experimentally in \cite{Combes2003a} and numerically in \cite{hinson2015}. The resultant model of the wing flapping dynamics is
\begin{equation}
\ddot\eta + \left(\Omega - (P^2+Q^2)I_{n_m}\right)\eta = M_a\left[\begin{array}{c} x_r\left(2PR - \dot{Q}\right) \\ -QR - \dot{P} \\ PR - \dot{Q} \end{array}\right],
\label{eqn:dynamics_no_aero}
\end{equation}
where $\Omega\in\mathbb{R}^{n_m\times n_m}$ is the diagonal modal frequency matrix, $M_a\in\mathbb{R}^{n_m\times 3}$ is the applied acceleration mass matrix, and $P,Q,R$ and $\dot P,\dot Q,\dot R$ are  wing rotation rates and wing rotational accelerations of the plate coordinate system, respectively.
Defining the state vector as $\mathbf{x}=\begin{bmatrix}\eta&\dot\eta&P&Q&R \end{bmatrix}^\top$ and the input vector as $\mathbf{u}=\begin{bmatrix}\dot P&\dot Q&\dot R\end{bmatrix}^\top$, the dynamics can be written in control-affine form as
\begin{equation}
    \begin{aligned}
        \dot{\vect{x}}&=\begin{bmatrix}\dot\eta\\ -K\eta+(2\mathbf{M_1}x_r+\mathbf{M_3})PR-\mathbf{M_2}QR\\0\\0\\0\end{bmatrix}+\begin{bmatrix}\mathbf{0}&\mathbf{0}&\mathbf{0}\\-\mathbf{M_2}&-\mathbf{M_1}x_r-\mathbf{M_3}&\mathbf{0}\\1&0&0\\0&1&0\\0&0&1\end{bmatrix}\mathbf{u},
    \end{aligned}\label{sys_wing}
\end{equation}
where $K=\Omega - (P^2+Q^2)I_{n_m}$ is the stiffness matrix and $\mathbf{M_j}\in\mathbb{R}^{n_m}$ denotes the $j$\textsuperscript{th} column of the matrix $M_a$.

\subsection{Measurements via Neural Encoding}
We review here the relevant neural encoding framework to relate strain due to the out-of-plane deformation, $\epsilon$, to the probability of firing of a neuron, $P_\text{fire}$, which we will assume to be the system output. A current, experimentally validated model of the neural response corresponding to strain stimulus, $\epsilon$, on a hawkmoth wing is based on two functions \cite{pratt2017}: the spike-triggered average (STA) and a nonlinear activation (NLA) function. The former is, as its name suggests, a functional approximation of the average of the stimuli which triggered an action potential at time $t=0$. As given below, it is approximated as an exponentially decaying sinusoidal function with a delay, $a$ \cite{mohren2018}, and its value is zeroed out as $t$ goes to infinity:
\begin{equation}\label{stave}
    \operatorname{STA}(t)=\cos{\left(\omega_\text{STA}(-t+a)\right)}\operatorname{exp}{\left(\frac{-(-t+a)^2}{b^2}\right)}.
\end{equation}
The other parameters, $\omega_\text{STA}$ and $b$, are the STA frequency in $\SI{}{\radian\per\second}$ and the width, respectively, and along with the delay parameter, $a$, they are taken to be constant for the neuron's encoding model.

\begin{figure}[hbt!] 
\centering
\includegraphics[width=0.7\textwidth]{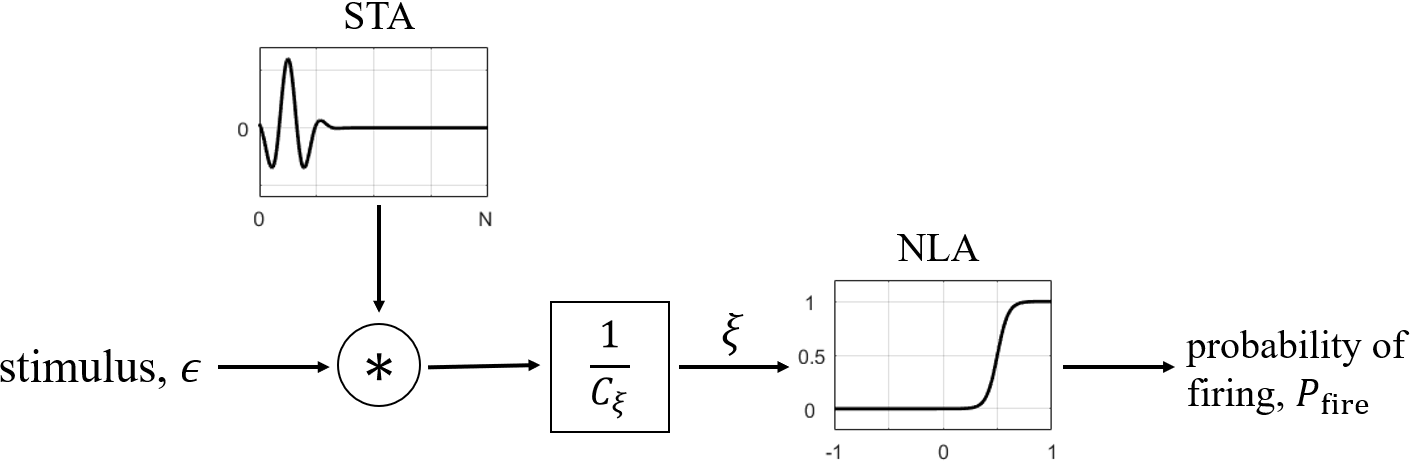}
\caption{The probabilistic firing model of a strain-sensitive neuron takes the strain data as input and calculates the probability of firing, $P_\text{fire}$, in two stages. First, the discrete convolution, denoted by $\ast$, of the strain data with $\text{STA}(\tau)$ is performed. After the normalization of filtered data by $1/C_\xi$, the probability of firing is given by $NLA(\xi)$.}\label{NEmodel}
\end{figure}

Define the projected stimulus as the convolution of a new strain stimulus, $\epsilon$, and the STA, that is,
\begin{align}\label{strain_proj}
    \xi(x_p,y_p,t)=\frac{1}{C_\xi}\int_{0}^{N}\epsilon(x_p,y_p,t-\tau) STA(\tau) d\tau,
\end{align}
where $N$ is the finite filter duration, and $C_\xi$ is a normalization constant which is further discussed in \cite{Boyacioglu2021}. This convolution can be used to estimate the firing rate of the neuron. However, the rate cannot be negative, and it also requires a saturation value to properly reflect the neuron's nonlinear behaviour \cite{dayan2001theoretical}. Hence the following nonlinear function of the projected stimulus is used to satisfy these requirements:
\begin{equation}\label{NLA}
\operatorname{NLA}(\xi)=\frac{1}{1+\exp (-c(\xi-d))}.
\end{equation}
Here, $c$ determines the slope and $d$ is the half-maximum position of the $\operatorname{NLA}$ function, and again, these parameters are treated as constants in the encoding model of a neuron. Finally, the probability of a neuron firing at the coordinates $(x_p,y_p)$ on the wing can be captured by the functional relationship \cite{mohren2018}:
\begin{equation}\label{probFire}
    P_{\text{fire}}(x_p,y_p,t) =NLA(\xi(x_p,y_p,t)),
\end{equation}
The full neural encoding model described above is illustrated in Fig. \ref{NEmodel}. Notice that the output is delayed due to the convolution of $\epsilon$ with the STA. It is also composed of two functions, $\xi$ and the NLA, which motivated us to study systems with these output types.

\section{Observability Background}\label{sec:back}

In this section, we review the relevant portions of observability that will be leveraged in the subsequent work.  The first approach is the common analytical observability for both linear and nonlinear systems.  We then provide an overview of the computational observability tool, the empirical observability Gramian, for systems where analytical methods are intractable.

\subsection{Analytical Observability} \label{anObsBack}
As a fundamental system property, observability characterizes the existence of an invertible mapping from measurements to state space. In this subsection, analytical tools for observability analysis of delay-free systems will be summarized along with the terminology used in the paper. Formally, a time-dependent system is {\bf\em observable} if there exists a finite time,  $T>0$, such that the knowledge of the input and the output over the time interval $[0,T]$ is sufficient to uniquely determine any unknown initial state \cite{Chen1999}. If the unknown initial state can be uniquely determined in an open neighborhood of the state, then the system is {\bf\em weakly observable.}

First, consider the continuous-time/discrete-time linear time-invariant control system
 \begin{equation}\label{wo_delay} 
  \Sigma_\text{CT-LTI}:\quad
  \begin{aligned}
{\dot{\vect{x}}}(t) &= A\vect{x}(t)+B\vect{u}(t)\\
   \vect{y}(t) &= C\vect{x}(t),
\end{aligned}
\quad / \quad  \Sigma_\text{DT-LTI}:\quad
  \begin{aligned}
{\vect{x}}_{k+1} &= A\vect{x}_k+B\vect{u}_k\\
   \vect{y}_k &= C\vect{x}_k,
\end{aligned}
\end{equation}
where $\vect{x}\in\mathbb{R}^n$ is the vector of states, $\vect{u}\in\mathbb{R}^m$ is the vector of inputs, and $\vect{y}\in\mathbb{R}^p$ is the vector of outputs. 
It is a well-known theorem that $\Sigma_\text{CT-LTI}$/$\Sigma_\text{DT-LTI}$ is observable if and only if the $T$-step observability matrix
\begin{equation}\label{observabilityk}
    \mathcal{O}_T=\begin{bmatrix}C\\ CA\\\vdots\\ CA^{T-1} \end{bmatrix}
\end{equation}
is full column rank for $T\geq n$ \cite{sontag1990}.
This observability rank condition implies that the input does not affect the observability of a linear system, which is not necessarily the case for a nonlinear system. 

Now, consider continuous-time nonlinear time-invariant systems in control-affine form:
 \begin{equation}\label{cont_affine}
 \Sigma_\text{NLTI}:\quad
   \begin{aligned}
     \dot{\vect{x}}(t) = \vect{f}_0(\vect{x}(t)) + \sum\limits_{i=1}^m \vect{f}_i(\vect{x}(t))u_i(t),\quad
     \vect{y}(t) = \vect{h}(\vect{x}(t)).
  \end{aligned}
 \end{equation}
The observability of such systems can be studied using differential geometry methods  \cite{Hermann1977,Anguelova2004}. To begin, one constructs the observability Lie algebra, $\mathcal{G}$, also termed the observation space,  for $\Sigma_\text{NLTI}$ \cite{Anguelova2004}:
\begin{equation}
    \mathcal{G}= \mathrm{span}\{L_{\vect{X}_1}\cdots L_{\vect{X}_k}\vect{h},\; k=0,1,\dots n-1, \; \vect{X}_k \in \{\vect{f}_0,\dots,\vect{f}_m\}\}
\end{equation}
where the Lie derivative, $L_{\vect{f}}\vect{h}$, is defined by
\begin{align}
     L_{\vect{f}} \vect{h}=\frac{\partial \vect{h}}{\partial \vect{x}}\vect{f},
\end{align}
and repeated derivatives are calculated as $L_{\vect{f}}^k \vect{h}=     L_{\vect{f}}L_{\vect{f}}^{k-1} \vect{h}$. A rank condition for weak observability is then
\begin{align}
    \mathrm{rank}(\mathrm{d}\mathcal{G})=n,
\end{align}
where $\mathrm{d}\mathcal{G}$ is the set of all gradients of the vector fields in $\mathcal{G}$, i.e. $\mathrm{d}\mathcal{G}=\{\mathrm{d}\phi:\phi\in\mathcal{G}\}$ \cite{Hermann1977}. Notice that because the control vector fields, $\vect{f}_i$, appear in the observability Lie algebra, the coupling of actuation with measurements can play a direct role in nonlinear observability, i.e., having a control term can make an unobservable control-free system observable.

\subsection{Observability Gramians and Output Sensitivity}
The observability rank conditions presented in the previous subsection provide a binary measure of observability and require the system to be expressed in certain forms. On the other hand, as shown below, quantitative measures of observability can be obtained via the analytical observability Gramian or the empirical observability Gramian matrix \cite{Krener2009} which, at the cost of some accuracy, has the benefit of not requiring analytical calculations such as the derivative of the measurement function, $\vect{h}(\vect{x})$, with respect to $\vect{x}$.

In continuous-time linear systems on the form of (\ref{wo_delay}), observability can also be determined from the nonnegative-definite symmetric matrix, $W_o(t_0,T)$, termed the {\it observability Gramian} \cite{Georges-IDRIM-2020} where
\begin{align}\label{linGram}
    W_o(t_0,T) &= \int_{t_0}^T e^{A^\top t}C^\top C e^{At} dt.
\end{align}
If this matrix has non-zero determinant (full rank), then the system is observable at time $t=t_0$.  In addition, the observability Gramian has the benefit of being a convex function to which a norm can be applied and a more refined measure of observability can be determined than the binary results above.  Of further interest is the relationship between $W_o(t_0,T)$ and the sensitivity of the output to the state being estimated, $\vect{x}(t_0)$.  Specifically, for linear time-invariant systems, one has the relation $\partial_{x_0} \vect{y}(t) = Ce^{At}$ which allows us to write 
\begin{align}
    W_o(t_0,T) & = \int_{t_0}^T \partial_{x_0} \vect{y}^\top(t) \partial_{x_0} \vect{y}(t) dt.
\end{align}
In this last form, the observability Gramian does not require the structure of a linear system, simply that the partial derivatives of the output with respect to the initial state can be computed \cite{Georges-IDRIM-2020}.

To implement the empirical approach, one simulates the system dynamics with perturbed initial conditions and an input sequence, $\vect{u}\in\mathcal{U}$, where $\mathcal{U}$ is the set of permissible controls.
Specifically, let $\vect{y}^{+i}$ and $\vect{y}^{-i}$ be the simulation outputs resulting from simulating the system dynamics with the nominal initial condition of state $\vect{x}_{0,i}$ perturbed respectively in the positive and negative directions by amount $\varepsilon$. Then the empirical observability Gramian is computed as
\begin{equation}
\label{eq:gramian}
W_o^\varepsilon(t_1, \mathbf{x}_0, \vect{u}) = \frac{1}{4\varepsilon^2} \int_0^{t_1} \Phi^\varepsilon(t,\mathbf{x}_0,\vect{u})^\top \Phi^\varepsilon(t,\mathbf{x}_0,\vect{u}) dt,
\end{equation}
where $\mathbf{x}_0$ is the initial state and
\begin{equation}
\label{eq:phi}
\Phi^\varepsilon(t,\mathbf{x}_0,\vect{u}) = \begin{bmatrix}\mathbf{y}^{+1}-\mathbf{y}^{-1} & \cdots & \mathbf{y}^{+n}-\mathbf{y}^{-n}\end{bmatrix}.
\end{equation}

In linear system observability analysis, the observability Gramian (\ref{linGram}) must be full rank for the system to be observable. The following theorem suggests a similar rank condition for $W_o^\varepsilon$.
 \begin{theorem}
\cite{Powel2015} Given $\Sigma_\text{NLTI}$, if there exists $\vect{u}\in\mathcal{U}$ such that
\begin{equation}
\rank\left(\lim_{\varepsilon \to 0} W_o^\varepsilon(t_1, \mathbf{x}_0, \vect{u})\right) = n
\end{equation}
for some $t_1>0$, then the system is weakly observable at $\mathbf{x}_0$.
\end{theorem}

\subsection{Observability Metrics}
In this subsection, we summarize observability metrics based on the (empirical) Gramian.
The two unobservability metrics introduced by Krener and Ide \cite{Krener2009} are the reciprocal of the minimum eigenvalue of the observability Gramian, $1/\ubar{\lambda}(W_o)$, which is also called the unobservability index, $\nu(W_o)$, and the condition number of the same matrix, $\kappa(W_o)= \bar{\lambda}(W_o)/\ubar{\lambda}(W_o)$. The former determines the weakness of the chain by its least strong link. The latter shows the balanced contribution of states to the output, and its value is desired to be one assuming that the output coordinates are already scaled.
It is a significant measure as it shows how well-conditioned the estimation problem is.

In \cite{Qi2014}, the optimal phasor measurement unit (PMU) placement problem was formulated to maximize the determinant of the observability Gramian, $\determ{W_o}$, which is equal to the product of all the eigenvalues of $W_o$, but it was advised that one should check the minimum eigenvalue to be at an acceptable level when using this approach.
Since $\determ{W_o}$ is not convex, $\log\determ{W_o}$ is sometimes preferred to be maximized instead, e.g. in \cite{Serpas2013}. However, the $n$-th root of the determinant, $[\determ{W_o}]^{1/n}$, is not only concave but also evaluates to zero when the system is unobservable, which is a necessary condition for an efficient metric \cite{muller1972}.

Although the trace and the spectral radius might be useful for applications like model reduction, they are not usually the first choice for observability analysis as they are not able to catch the case when $W_o$ is singular.

\section{Observability Analysis of Systems with Output Delay}\label{sec:delay}
In this section, we first present our approach to the observability analysis of linear systems with output delay and by relating the observability matrices of the cases of delay-free and delayed measurements show that having a filter can make an observable system unobservable, but not vice versa. Second, we use the existing geometrical methods to study the related observability of nonlinear autonomous systems.

In real systems, all measurements are delayed as compared to the standard non-delayed model. When the output is an explicit function of a state vector at a previous time, it is said the system has an ideal delay. Whereas here we consider systems with an output function explicitly depending on more than one previous state vectors.

\subsection{Linear Systems with Delay}
Consider the following discrete-time time-invariant system with linear dynamics and output delay:
\begin{equation}
 \Sigma_\text{LTIw/D}:\quad
    \begin{aligned}\label{sys_delay}
        \vect{x}_{k+1}&=A\vect{x}_k+B\vect{u}_k,\quad
        \vect{y}_k&=\sum_{\tau=0}^NC_\tau\vect{x}_{k-\tau}.
    \end{aligned}
\end{equation}
Here, $\vect{x}_{k-N}$ is the oldest state upon which $\vect{y}_k$ explicitly depends.
 \begin{corollary} \label{cor1}The system with delay, $\Sigma_\text{LTIw/D}$, is observable if and only if the $p n\times n$ observability matrix
\begin{equation}\label{obs_barC}
    \mathcal{\bar O}=\begin{bmatrix} \bar C\\\bar CA\\\vdots\\\bar CA^{n-1}\end{bmatrix},
\end{equation}
is full column rank where the $p\times n$ matrix, $\bar C$, is defined as:
\begin{equation}
    \bar C=\sum_{\tau=0}^NC_\tau A^{N-\tau}.
\end{equation}

\begin{Proof} First consider the output at time $k=0$, $\vect{y}_0$, and use the system dynamics to relate it to only  $\vect{x}_{-N}$:
\begin{align*}
    \vect{y}_0&=C_0\vect{x}_0+C_1\vect{x}_{-1}+\dots+C_N\vect{x}_{-N}\\
        &=C_0(A\vect{x}_{-1}+B\vect{u}_{-1})+C_1\vect{x}_{-1}+C_2\vect{x}_{-2}+\dots+C_N\vect{x}_{-N}\\
    &~~\vdots\\
    &=C_0B\vect{u}_{-1}+(C_0A+C_1)B\vect{u}_{-2}+\dots+(C_0A^{N-1}+\dots+C_{N-2}A+C_{N-1})B\vect{u}_{-N}+(C_0A^N+\dots+C_{N-1}A+C_N)\vect{x}_{-N}\\
    &=\sum_{\tau=1}^N \left(\sum_{i=1}^\tau C_{i-1}A^{\tau-i}\right)B\vect{u}_{-\tau}+\sum_{\tau=0}^NC_\tau A^{N-\tau}\vect{x}_{-N}
\end{align*}
Hence, a set of $n$ consecutive measurements can be constructed as follows and as is usually done for delay-free systems:
\begin{equation*}
    \begin{aligned}
    \underline{\vect{y}}_k&=\bar C\vect{x}_{k-N}\\
    \underline{\vect{y}}_{k+1}&=\bar C\vect{x}_{k-N+1}=\bar CA\vect{x}_{k-N}+\bar CB\vect{u}_{k-N}\\
    &~~\vdots\\
    \underline{\vect{y}}_{k+n-1}&=
    \bar C\vect{x}_{k-N+n-1}=\bar CA^{n-1}\vect{x}_{k-N}+\sum_{j=0}^{n-2}\bar CA^{n-2-j}B\vect{u}_{k-N+j}
    \end{aligned}
\end{equation*}
where
\begin{equation}
    \underline{\vect{y}}_k=\vect{y}_k-\sum_{\tau=1}^N \left(\sum_{i=1}^\tau C_{i-1}A^{\tau-i}\right)B\vect{u}_{k-\tau}.
\end{equation}
Finally, we can write
\begin{align}
    \doubleunderline{\vect{y}}=\mathcal{\bar O}\vect{x}_{k-N},
\end{align}
where the $pn\times1$ vector $\doubleunderline{\vect{y}}$ is defined as
\begin{align}
    \doubleunderline{\vect{y}}=\begin{bmatrix} \underline{\vect{y}}_k\\\underline{\vect{y}}_{k+1}\\\vdots\\\vdots\\\underline{\vect{y}}_{k+n-1}\end{bmatrix}-\begin{bmatrix}\vect{0}&\vect{0}&\cdots&\vect{0}\\\bar CB&\vect{0}&\ddots&\vect{0}\\\vdots&\ddots&\vdots&\vdots\\\vdots&\vdots&\ddots&\vdots\\\bar CA^{n-2}B&\bar CA^{n-3}B&\cdots& \bar CB \end{bmatrix}\begin{bmatrix}\vect{u}_{k-N}\\\vect{u}_{k-N+1}\\\vdots\\\vect{u}_{k-N+n-2} \end{bmatrix}.
\end{align}
Here, $\doubleunderline{\vect{y}}$ is known, and $\vect{x}_{k-N}$ can be uniquely determined if and only if the nullity of the observability matrix is zero.
\hfill$\blacksquare$
\end{Proof}
\end{corollary}

Corollary (\ref{cor1}) also implies that the observability of linear systems with or without delay does not depend on the input.

In the rest of this section, two special cases will be considered. In the first case, the same filter coefficients are used for each output, $(\vect{y}_k)_i$, and the coefficient matrices are assumed to be a scaled version of the measurement matrix, $C$, of Eq. (\ref{wo_delay}). In the case of heterogeneous sensing, the coefficient matrices are obtained by scaling each row of $C$ separately. It will be shown that in both of these noteworthy cases, having an output delay does not make an unobservable delay-free linear system observable. This result is an expected one because the methods we use here are just an application of well-known types of filtering.
\subsubsection{Uniform Sensing}
If $C_\tau=\gamma_\tau C$ in Eq. (\ref{sys_delay}) where $\gamma_\tau\in\mathbb{R}$ is a scalar, then the observability matrix (\ref{obs_barC}) becomes:
\begin{align*}
    \mathcal{\bar O}&=\begin{bmatrix} \gamma_NC+\gamma_{N-1}CA+\cdots+\gamma_0CA^N\\
    \gamma_NCA+\gamma_{N-1}CA^2+\cdots+\gamma_0CA^{N+1}\\
    \vdots\\
    \gamma_NCA^{n-1}+\gamma_{N-1}CA^n+\cdots+\gamma_0CA^{N+n-1}
    \end{bmatrix}.
\end{align*}

\begin{corollary}\label{cor:linUniform}
If the delay-free system, $\Sigma_\text{DT-LTI}$, is unobservable, then $\Sigma_\text{LTIw/D}$ with uniform sensing is also unobservable. Also, if the matrix polynomial $(I\gamma_N+A\gamma_{N-1}+\cdots+A^N\gamma_0)$ is nonsingular, then $\rank(\mathcal{\bar O})=\rank(\mathcal{O}_n)$.
\begin{Proof}
The observability matrix $\mathcal{\bar O}$ can be written as
\begin{align}\label{matpol}
\mathcal{\bar O}=\gamma_N\begin{bmatrix}C\\ CA\\\vdots\\ CA^{n-1} \end{bmatrix}+\gamma_{N-1}\begin{bmatrix}CA\\ CA^2\\\vdots\\ CA^{n} \end{bmatrix}+\cdots+\gamma_{0}\begin{bmatrix}CA^N\\ CA^{N+1}\\\vdots\\ CA^{N+n-1} \end{bmatrix}=\mathcal{O}_n(I_n\gamma_N+A\gamma_{N-1}+\cdots+A^N\gamma_0),\end{align}
and the result follows directly.\hfill$\blacksquare$
\end{Proof}
\end{corollary}

The exceptional case of the singularity of the matrix polynomial is illustrated in \cite{Boyacioglu2021}.
 
\subsubsection{Heterogeneous Sensing}
If $C_\tau=\diag{(\gamma_{1\tau},\gamma_{2\tau},\dots,\gamma_{p\tau})}C=G_\tau C$ in Eq. (\ref{sys_delay}), then the observability matrix (\ref{obs_barC}) becomes
\begin{align*}
    \mathcal{\bar O}&=\begin{bmatrix} G_NC+G_{N-1}CA+\cdots+G_0CA^N\\
    G_NCA+G_{N-1}CA^2+\cdots+G_0CA^{N+1}\\
    \vdots\\
    G_NCA^{n-1}+G_{N-1}CA^n+\cdots+G_0CA^{N+n-1}
    \end{bmatrix}=\begin{bmatrix} G_N&G_{N-1}&\cdots&G_0&0&\cdots&0\\
    0& G_N&G_{N-1}&\cdots&G_0&\cdots&0\\
    \vdots&\ddots&\ddots&\ddots&\ddots&\ddots&\vdots\\
    0&\cdots&0&G_N&G_{N-1}&\cdots&G_0
    \end{bmatrix} \begin{bmatrix} C\\CA\\\vdots\\CA^{N+n-1}\end{bmatrix}.
\end{align*}
\begin{corollary}
If the delay-free system, $\Sigma_\text{DT-LTI}$, is unobservable, then $\Sigma_\text{LTIw/D}$ with heterogeneous sensing is also unobservable.

\begin{Proof}
Since $\rank(\mathcal{O}_{N+n})=\rank(\mathcal{O}_n)$, and the rank of a matrix product is no larger than the minimum of the ranks of the factors, it can be stated that $\rank{(\mathcal{\bar O})}\leq\rank(\mathcal{O}_n)$.\hfill$\blacksquare$
\end{Proof}\end{corollary}

Note that this analysis is also valid where $G_\tau$ is not a diagonal matrix.

Systems where each output has a window size are not within the scope of this study. However, an approach for the analysis of such systems is outlined as follows. Consider the case where $ (\vect{y}_k)_i=\sum_{\tau=0}^{N_i}\gamma_{i\tau}C\vect{x}_{k-\tau}$. Then define $N=\max{(N_1,N_2,\dots,N_p)}$, and construct output matrices as $C_\tau=\diag{(\gamma_{1\tau},\gamma_{2\tau},\dots,\gamma_{p\tau})}C$ by letting $\gamma_{i\tau}=0$ if it is not already defined. Finally, a sufficient observability condition can be obtained by following similar steps to the constant-history-length heterogeneous sensing case.

\subsection{Nonlinear Systems with Delay} \label{sec:nonlinear}
In this section, we describe how the existing geometrical methods presented in Section \ref{sec:back} can be implemented for nonlinear systems with delay. Our analytical approach is restricted to autonomous systems where $\mathbf{u}=\vect{0}$ because of the constraints below. The non-autonomous system can be studied using the empirical observability Gramian.

Consider the following class of autonomous systems with output delay:
\begin{equation}\label{sys_long}
 \Sigma_\text{NLTIw/D}:\quad
    \begin{aligned}
        \dot{\vect{x}}(t) &= \vect{f}_0(\vect{x}(t)),\quad
        y_i(t) &= \int_{0}^{N}C(\tau)x_j(t-\tau)d\tau,\quad i=1,\dots,p,~j\in \{1,\dots,n\}
    \end{aligned}
\end{equation}
Here, each of the $p$ outputs is the convolution of one of the states with the same coefficient function, $C$. Assume that all states appearing in the output vector can be obtained as a function of $\vect{x}(t-N)$ and $\tau$, that is, assume that there exists a vector function, $\vect{\bar h_{\text{aux}}}$, such that
\begin{align}
    x_j(t-\tau)=\bar h_{\text{aux}_i}(\vect{x}(t-N),\tau).
\end{align}
Since it is possible to write the output as a function of only $\vect{x}(t-N)$, the Lie algebraic approach can be used for the observability analysis as is done for delay-free nonlinear systems. The observation space, $\mathcal{\bar G}$, for $\Sigma_\text{NLTIw/D}$ is obtained by
\begin{equation}
    \mathcal{\bar G}= \mathrm{span}\{L_{\vect{f}_0}^{n-1} \mathbf{\bar h},\dots,L_{\vect{f}_0}\mathbf{\bar h},\mathbf{\bar h}\},
\end{equation}
where the Lie derivative is defined by
\begin{align}
     L_{\vect{f}_0} \mathbf{\bar h}=\frac{\partial \mathbf{\bar h}(\vect{x}(t-N))}{\partial \vect{x}(t-N)}{\vect{f}_0},
\end{align}
and $\mathbf{\bar h}$ is defined in
\begin{align}\label{def:hbar}
    \vect{y}(t)=\mathbf{\bar h}(\vect{x}(t-N))=   \int_{0}^{N}C(\tau)\vect{\bar h_{\text{aux}}}(\vect{x}(t-N),\tau)d\tau.
\end{align}
Here, obtaining the set of all gradients of the vector fields in $\mathcal{\bar G}$, $\mathrm{d}\mathcal{\bar G}=\{\mathrm{d}\phi:\phi\in\mathcal{\bar G}\}$, requires solving the integral in Eq. (\ref{def:hbar}) which is usually a cumbersome process. The following proposition is offered to sometimes bypass this process.
\begin{Proposition}\label{prop:rank}
Define $\mathcal{\bar G}_{\text{aux}}= \mathrm{span}\{L_{\vect{f}_0}^{n-1} \mathbf{\bar h_{\text{aux}}},\dots,L_{\vect{f}_0}\mathbf{ \bar h_{\text{aux}}},\mathbf{ \bar h_\text{aux}}\}$ and $\mathrm{d}\mathcal{\bar G}_{\text{aux}}=\{\mathrm{d}\phi:\phi\in\mathcal{\bar G}_{\text{aux}}\}$. If the rows/columns of $d\mathcal{\bar G}_{\text{aux}}$ are linearly dependent, then $\mathrm{d}\mathcal{\bar G}$ is rank-deficient.
\begin{Proof}
Let $\rank{(d\mathcal{\bar G}_{\text{aux}})}<n$. Then the Leibniz integral rule allows us to write
\begin{align}\label{eq:inner}
    d\mathcal{\bar G}&=\int_{0}^{N}C(\tau)d\mathcal{\bar G}_{\text{aux}}d\tau.
\end{align}
If the columns/rows of $d\mathcal{\bar G}_{\text{aux}}$ are linearly dependent, then $\rank{(d\mathcal{\bar G})}<n$. \hfill$\blacksquare$
\end{Proof}
\end{Proposition}
\subsection{Example Systems}
In \cite{Boyacioglu2021}, we referred to Proposition \ref{prop:rank} to show that flexible wing flapping dynamics (\ref{sys_wing}) with filtered output and without control ($\mathbf{u}=\vect{0}$) are unobservable.
Here, we give an example to illustrate the observability analysis of a linear system with output delay.
\subsubsection{Double Integrator Dynamics with Differencing Output}
Consider the discretized dynamics of a double integrator with the sampling period, $T_s$, and differencing output:
  \begin{equation}\label{sys_outDiff}
  \begin{split}
   \vect{x}_{k+1} = \begin{bmatrix}1&T_s\\0&1 \end{bmatrix}\vect{x}_k,\qquad
   y_k = \begin{bmatrix}1&0 \end{bmatrix}\vect{x}_k-\begin{bmatrix}1&0 \end{bmatrix}\vect{x}_{k-1}.\\
  \end{split}
 \end{equation} Here, we obtain $\bar C$ as
\begin{equation*}
    \bar C=\begin{bmatrix}1&0 \end{bmatrix}\begin{bmatrix}1&T_s\\0&1 \end{bmatrix}-\begin{bmatrix}1&0 \end{bmatrix}=\begin{bmatrix}
    0&T_s
    \end{bmatrix},
\end{equation*}
and the observability matrix becomes
\begin{equation*}
    \bar O=\begin{bmatrix}
    0&T_s\\0&T_s
    \end{bmatrix}
\end{equation*}
which is rank-deficient for any $T_s$. As Cor. \ref{cor:linUniform} suggests, this result could also be obtained by showing that
\begin{equation*}
    I_n\gamma_1 +A\gamma_{0}=I_2-\begin{bmatrix}
    1 &T_s\\0 &1
    \end{bmatrix}
\end{equation*} is singular for any $T_s$. That is to say, the double integrator dynamics with differencing output (\ref{sys_outDiff}) is unobservable, although the same system with the output $y_k = \begin{bmatrix}1&0 \end{bmatrix}\vect{x}_k$ is observable.

\section{Observability Analysis of Systems with a Composite Output Function}\label{sec:compost}
In this section, we present theoretical results on the observability of systems with an output having the structure of function composition of the form, $g \circ h$. Our first main results is that if a system
\begin{equation}
\Sigma_h:\quad
\begin{aligned}
\dot{\vect{x}}(t) = \vect{f}_0(\vect{x}(t)),\quad
y(t)  = h(\vect{x}(t)),
\end{aligned}
\end{equation}
is unobservable, then the system
\begin{equation}\label{sys_goh}
\Sigma_{g\circ h}:\quad
\begin{aligned}
\dot{\vect{x}}(t) = \vect{f}_0(\vect{x}(t)),\quad
y(t)  = (g\circ h)(\vect{x}(t)),
\end{aligned}
\end{equation}
is also unobservable for any function $g$. Applying a function $g$ to the output of an unobservable system thus cannot make the unobservable system observable. Our second main result is that $\Sigma_{g\circ h}$ can be unobservable, even if $\Sigma_h$ is observable. When applying a function $g$ to the output of $\Sigma_h$ leads to an unobservable system $\Sigma_{g\circ h}$, we term the combination of $\Sigma_h$ and $g$ a singular case. One should avoid singular cases whenever possible. For the remainder of this section, we consider cases in which $g$ and $h$ have the same output-space dimension.

\subsection{Observability Space for Systems with Composite Output Functions}
We first consider a control-free nonlinear system $\Sigma_{h}$ with a single output, i.e., $\vect{f}_0:\mathbb{R}^n\rightarrow \mathbb{R}^n$ and $h:\mathbb{R}^n\rightarrow \mathbb{R}$. We also consider another system $\Sigma_{g\circ h}$, which has the same internal dynamics as the first system but uses a function composition $g\circ h$ as output, where $g:\mathbb{R}\rightarrow \mathbb{R}$.

The two systems have the respective observation spaces
\begin{align}
    \mathcal{G}_h&=\mathrm{span}\{L_{\vect{f}_0}^{n-1} h,\dots,L_{\vect{f}_0}h,h\}\label{liealgebra1}\\
    \mathcal{G}_{g\circ h}&=\mathrm{span}\{L_{\vect{f}_0}^{n-1} (g\circ h),\dots,L_{\vect{f}_0}(g\circ h),g\circ h\}\label{liealgebra2}
\end{align}
Using the gradient operator, $\nabla$, where
    \begin{align}
        \nabla\phi := \left(
            \begin{matrix}
                \frac{\partial \phi}{\partial x_1}&
                \frac{\partial \phi}{\partial x_2}&
                \cdots&
                \frac{\partial \phi}{\partial x_n}
            \end{matrix}
            \right)^\top,
    \end{align}
one can obtain the Jacobian matrices 
\begin{gather}
    d\mathcal{G}_h=
    \begin{bmatrix}\nabla h&\nabla L_\mathbf{f_0}h&\cdots&\nabla L^{n-1}_\mathbf{f_0}h \end{bmatrix}^\top,\\
    d\mathcal{G}_{g\circ h}=
    \begin{bmatrix}\nabla (g\circ h)&\nabla L_\mathbf{f_0}(g\circ h)&\cdots&\nabla L^{n-1}_\mathbf{f_0}(g\circ h) \end{bmatrix}^\top\label{dGgoh},
\end{gather}
of the observation spaces. 
Each of the Jacobian matrices has full rank if and only if the corresponding system is observable. We thus call $d\mathcal{G}_h$ and $d\mathcal{G}_{g\circ h}$ the \textit{observability matrices} of $\Sigma_{h}$ and $\Sigma_{g\circ h}$, respectively.
\subsection{Higher-order Lie Derivatives of Composite Functions}
To be able to relate the observability of $\Sigma_{g\circ h}$ to $\Sigma_{h}$, we are interested in the determinant of the two observability matrices. $d\mathcal{G}_{g\circ h}$ and $d\mathcal{G}_h$. We start with deriving an expression for
$L^k_{\bf f_0}(g\circ h)$ which appears in Eq. (\ref{liealgebra2}) and its gradient appears in Eq. (\ref{dGgoh}) where $k$ is a positive integer.
\begin{Lemma}
The $k$-th Lie derivative of a composition $g\circ h$ with respect to a vector field $\vect{f}$ is
\begin{align}
    L^k_{\bf f}(g\circ h)& = \sum_{j=1}^k\left[(g^{(j)}\circ h) \sum_{s\in M_{k,j}}\left(\prod_{s_i\in s} L_{\bf f}^{s_i}(h)\right)\right]\,,\label{eq:lemma}
\end{align}
where $g^{(j)}\circ h$ denotes the $j$-th derivative of $g$ with respect to $h$ and $M_{k,j}$ is a multiset of ordered multisets of integers with the recursive construction rule
\begin{align}
    M_{k,j} := 
        \left\{
        \begin{matrix}
        \left\{s\backslash \{s_i\} \cup \{s_{i+1}\}\:|\:s_i\in s,\:s\in M_{k-1,j} \right\} \cup
        \left\{s \cup \{1\}\:|\:s\in M_{k-1,j-1}\right\} & \textrm{if $k>j>0$,}\\
        \emptyset & \textrm{otherwise.}
        \end{matrix}
        \right.\label{eq:mdef1}
\end{align}

\begin{Proof}
See the Appendix.\hfill$\blacksquare$
\end{Proof}
\end{Lemma}

\begin{corollary}\label{cor:mainLie}
The gradient of the $k$-th Lie derivative of a composition $g\circ h$ is
\begin{align}
    \nabla L^k_{\bf f}(g\circ h)& = \sum_{j=1}^k\left\{(g^{(j+1)}\circ h)\nabla h \sum_{s\in M_{k,j}}\left[\prod_{s_i\in s} L_{\bf f}^{s_i}(h)\right]+(g^{(j)}\circ h)\sum_{s\in M_{k,j}}\sum_{s_\ell\in s}\left[\nabla\left(L_{\bf f}^{s_\ell}(h)\right)\prod_{\substack{s_i\in s\,,\\i\neq\ell}} L_{\bf f}^{s_i}(h)\right]
    \right\}\,.\label{eq:coro}
\end{align}

\begin{Proof}
We apply the product rule for derivatives to Eq.\,(\ref{eq:lemma}) and obtain
\begin{align}
    \nabla L^k_{\bf f}(g\circ h)& = \sum_{j=1}^k\left\{(g^{(j+1)}\circ h)\nabla h \sum_{s\in M_{k,j}}\left[\prod_{s_i\in s} L_{\bf f}^{s_i}(h)\right]+(g^{(j)}\circ h)\sum_{s\in M_{k,j}}\left[\nabla\prod_{\substack{s_i\in s}} L_{\bf f}^{s_i}(h)\right]
    \right\}\,.\label{eq:proof_coro1}
\end{align}
Applying the product rule again to the second term in Eq.\,(\ref{eq:proof_coro1}) yields Eq.\,(\ref{eq:coro}).\hfill$\blacksquare$

\end{Proof}
\end{corollary}

\subsection{Observability of Systems with Composite Output Functions}
We are now ready to present the main result of this section by relating $d\mathcal{G}_{h}$ and $d\mathcal{G}_{g\circ h}$ using Cor. \ref{cor:mainLie}.

\begin{theorem}\label{theo:mainDet}
Given systems $\Sigma_h$ in (32) and $\Sigma_{g \circ h}$ in (33), define the observation spaces $\mathcal{G}_{h}=\mathrm{span}\{L_{\vect{f}_0}^{n-1} h,\dots,L_{\vect{f}_0}h,h\}$ and $
    \mathcal{G}_{g\circ h}=\mathrm{span}\{L_{\vect{f}_0}^{n-1} (g\circ h),\dots,L_{\vect{f}_0}(g\circ h),g\circ h\}$ where $\vect{f}_0:\mathbb{R}^n\rightarrow \mathbb{R}^n$, $h:\mathbb{R}^n\rightarrow \mathbb{R}$, and $g:\mathbb{R}\rightarrow \mathbb{R}$. Then, the Jacobian matrices of $\mathcal{G}_1$ and $\mathcal{G}_2$ are related with the expression
\begin{equation}\label{eq:dets1}
    \det d\mathcal{G}_{g\circ h}=\left(\frac{dg}{dh}\right)^n\det d\mathcal{G}_{h}.
\end{equation}

\begin{Proof} Using (\ref{eq:coro}), it can be written that
\begin{equation*}\begin{aligned}
\mathcal{G}_{g\circ h}&=\begin{bmatrix} 
(g^{(1)}\circ h)\nabla^\top h\\\left\{
(g^{(2)}\circ h)\nabla h \sum_{s\in M_{1,1}}\left[\prod_{s_i\in s} L_{\bf f_0}^{s_i}(h)\right]+(g^{(1)}\circ h)\sum_{s\in M_{1,1}}\sum_{s_\ell\in s}\left[\nabla\left(L_{\bf f_0}^{s_\ell}(h)\right)\prod_{\substack{s_i\in s\,\\{i\neq\ell}}} L_{\bf f_0}^{s_i}(h)\right]\right\}^\top
\\\vdots\\
\sum_{j=1}^{n-1}\left\{(g^{(j+1)}\circ h)\nabla h \sum_{s\in M_{n-1,j}}\left[\prod_{s_i\in s} L_{\bf f_0}^{s_i}(h)\right]+(g^{(j)}\circ h)\sum_{s\in M_{n-1,j}}\sum_{s_\ell\in s}\left[\nabla\left(L_{\bf f_0}^{s_\ell}(h)\right)\prod_{\substack{s_i\in s\,\\{i\neq\ell}}} L_{\bf f_0}^{s_i}(h)\right]\right\}^\top
\end{bmatrix}.
\end{aligned}\end{equation*}
Since we are interested in the determinant of the Jacobian matrix, $\mathcal{G}_{g\circ h}$, we are able to use the Gaussian elimination algorithm and obtain
\begin{equation}
    \det{\mathcal{G}_{g\circ h}}=\begin{vmatrix} 
(g'\circ h)\nabla h&
(g'\circ h)\nabla L_{\bf f_0}h&\cdots&
(g'\circ h)\nabla L^{n-1}_{\bf f_0}h
\end{vmatrix}=(g'\circ h)^n\begin{vmatrix} 
\nabla h&
\nabla L_{\bf f_0}h&\cdots&
\nabla L^{n-1}_{\bf f_0}h
\end{vmatrix}=\left(\frac{dg}{dh}\right)^n\det{\mathcal{G}_{h}}.\end{equation}\hfill$\blacksquare$
\end{Proof}
\end{theorem}

We draw two implications from Theorem \ref{theo:mainDet}: \begin{enumerate}
    \item If the nonlinear system $\Sigma_{h}$
is unobservable, then the system $\Sigma_{g\circ h}$ with the same internal dynamics and composite output function $g\circ h$ is also unobservable. 
\item Observable systems $\Sigma_{h}$ lead to observable systems $\Sigma_{g\circ h}$ if and only if $dg/dh$ is non-zero.
\end{enumerate}

This result allows us to comment on the effect of $g$ on observability even if it is not possible to analytically study expressions like $\nabla h$. The examples given in the subsection below illustrate these aspects.

The observability analysis of controlled nonlinear systems requires  calculating Lie derivatives with respect to the control fields since the separation principle does not hold as shown in Sec. \ref{anObsBack}. The extension of the result presented for control-free systems to single- or multi-input systems requires an expression for $\nabla L^{n-1}_\mathbf{f_0}L^{n-1}_\mathbf{f_i}(g\circ h)$ similar to the one we obtained in Corollary \ref{cor:mainLie}. We anticipate that  the requirements for the observability of a system with a composite output function are the same for control-free systems and for single- or multiple-input systems with one difference: The observability matrix will not be square. Similarly, we would not have a square observability matrix for a system with multiple outputs.

\subsection{Example Systems}
In this section, we present two examples to illustrate the results from above for systems with an output function in the form of function composition.

\subsubsection{Double Integrator Dynamics with a Saturated Output}
First, consider the dynamics of a double integrator with an unsaturated output:
  \begin{equation}\label{sys_unsat}
  \begin{split}\Sigma_h:\quad
   \dot{\vect{x}}(t) = \begin{bmatrix}0&1\\0&0 \end{bmatrix}\vect{x}(t),\qquad
   y(t) = \begin{bmatrix}1&0 \end{bmatrix}\vect{x}(t).\\
  \end{split}
 \end{equation}
 Since the system is linear, the observability matrix can be easily written as
 $d\mathcal{G}_{h}=\diag{(1,1)}$ which is full rank, that is, the system (\ref{sys_unsat}) observable.

Now consider the same system dynamics with a saturated output:
  \begin{equation}\label{sys_sat}
  \begin{split}
   \Sigma_{g\circ h}:\quad\dot{\vect{x}}(t) = \begin{bmatrix}0&1\\0&0 \end{bmatrix}\vect{x}(t),\qquad
   y(t) = \tanh\left(\begin{bmatrix}1&0 \end{bmatrix}\vect{x}(t)\right),\\
  \end{split}
 \end{equation}
where the hyperbolic tangent function, which is infinitely differentiable, creates the saturation effect (Fig. \ref{hyptan}a). Since $d\mathcal{G}_{h}$ is full rank, by the help of Theorem \ref{theo:mainDet}, one can state that the system is observable for $[d(\tanh{x_1})/d x_1]^2=\sech^4{x_1}\neq0$. Indeed, the observation space of the system (\ref{sys_sat}) is $\mathcal{G}_{g\circ h}=\mathrm{span}\{L_{\vect{f}_0}(\tanh{x_1}),\tanh{x_1}\}$,
where $\mathbf{f_0}=\begin{bmatrix} x_2&0\end{bmatrix}^\top$, and the nonlinear observability matrix can be written as:
\begin{align*}
    d\mathcal{G}_{g\circ h}=\begin{bmatrix} \frac{\partial \tanh{x_1}}{\partial x_1}&\frac{\partial \tanh{x_1}}{\partial x_2}\\\frac{\partial (x_2\sech^2{x_1})}{\partial x_1}&\frac{\partial( x_2\sech^2{x_1})}{\partial x_2}\end{bmatrix}=\begin{bmatrix} \sech^2{x_1}&0\\-2x_2 \sech^2(x_1) \tanh(x_2)&\sech^2{x_1}\end{bmatrix},
\end{align*}
which is full rank for $\det{d\mathcal{G}_{g\circ h}}=\sech^4{x_1}\neq0$.

When the derivative of $\tanh{x_1}$ is practically zero, the initial state becomes indistinguishable. Similarly, when the slope is steep, it is easier to distinguish neighboring initial states, that is, when the derivative is large, we get higher output sensitivity as well as higher observability metrics such as the $n$-th root of the determinant of the observability Gramian. Finally, note that using a rectified hyperbolic tangent function is an alternative way to express the activation function in the neural encoding process \cite{dayan2001theoretical}.

\subsubsection{NLA Parameters Selection}
\begin{figure}[hbt!] 
\centering
\includegraphics[width=0.8\textwidth]{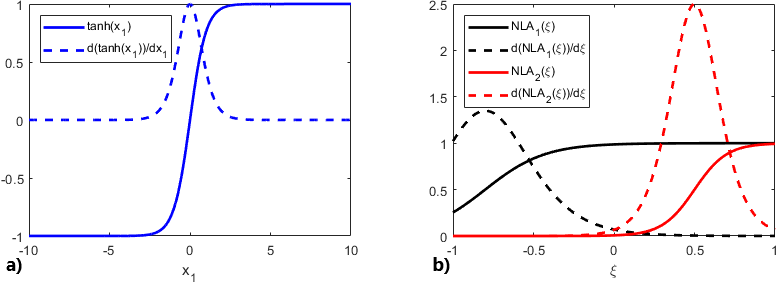}
\caption{The outer functions and their derivatives appeared in a) the first and b) second examples. Here, $\operatorname{NLA}(\xi)=1/1+\exp (-5.4(\xi+0.8))$ and $\operatorname{NLA}(\xi)=1/1+\exp (-10(\xi-0.5))$.
} \label{hyptan}
\end{figure}

Recall the nonlinear activation function on the projected stimulus, $\xi$, in the neural encoding process which gives the probability of firing of a neuron and is considered as the output of the system:
\begin{equation*}
\operatorname{NLA}(\xi)=\frac{1}{1+\exp (-c(\xi-d))}.
\end{equation*}
We are interested in the parameters of NLA, $c$ and $d$, and desire them not to make the system unobservable assuming that it is observable when the output is $y(t)=\xi(t)$. Note that this assumption of observability is based on the calculation of the approximate lower bounds for the empirical Gramian eigenvalues \cite{Boyacioglu2021}. Here, the role of the NLA function is the same as $g$ in (\ref{sys_goh}), and its derivative with respect to $\xi$ is
\begin{align}
    \frac{\operatorname{d} \operatorname{NLA}(\xi)}{\operatorname{d} \xi} =\frac{c\exp(-c(\xi-d))}{(\exp(-c(\xi-d))+1)^2}.
\end{align}
If there is no slope ($c=0$), the NLA function makes the system unobservable. Also, the derivative, and hence the output sensitivity, is expected to be higher about the axis $\xi=d$. In Fig. \ref{hyptan}b, two NLA functions and derivatives are plotted. One can imply that the first set of NLA parameters would serve better to observe the system when $\xi<0$ and the second is good to use when $\xi$ is positive.
Finally, note that due to the relation (\ref{eq:dets1}), we are able to assess the observability, even though the functional structure does not allow us to study it analytically.

\section{Hawkmoth Wing Numerical Application} \label{sec:sims}
In this section, we address some applications of interest with respect to biological sensing in the hawkmoth. Due to the complexity of the wing model dynamics and unavailable function derivatives, the empirical Gramian is used instead of the analytical Gramian. After the numerical observability analysis, we optimally solve the problem of neural-inspired sensor placement for the hawkmoth flapping wing model using two sensor types. Second, we present the effect of NLA parameters on the observability of flapping wing dynamics using a metric based on the empirical Gramian and compare the results with the ones that come from the derivative analysis in the previous section.

\subsection{System Simulations}
In order to implement the empirical Gramian, system simulations must be set up. The simulation of the full dynamics of a \textit{Manduca sexta} wing requires several components. First, the wing stroke kinematics, the effective inputs to the system, $\vect{u}$, are controlled by three Euler angles which provide the nominal input sequence. The planform geometry of the wing is based on photographs of a hawkmoth forewing from \cite{dickerson2014} and is scaled for a $\SI{50}{\milli\meter}$ wingspan. Finally, the wing structural mode shapes and frequencies and the wing mass density and thickness are taken from \cite{hinson2015}.

The following stroke kinematics were used to define a nominal flapping cycle:
\begin{align}
\theta(t) = 0,\quad
 \psi(t) = -A_\psi \cos(2\pi t/T_\text{beat}),\quad
 \alpha(t) = \frac{\pi}{2} - A_\alpha \tanh\left( \frac{\pi}{2} \sin(2\pi t/T_\text{beat})\right) .
\end{align}
Here, the stroke position amplitude, $A_\psi$, and the feathering angle amplitude, $A_\alpha$, are taken as $\ang{45} $ and $60$, respectively, and the wing-beat period, $T_\text{beat}$, is $\SI{40}{\milli\second}$. The simulation time is $2T_\text{beat}$, where the first period is used to provide the stimulus history required for the filter. The trajectory is perturbed at the beginning of the second period, and the empirical observability Gramian is calculated for that second period, that is, $t_1$ in Eq. (\ref{eq:gramian}) is chosen to be $T_\text{beat}$.

The STA and NLA parameters were selected based on values derived in \cite{mohren2018} using electrophysiological recordings of mechanoreceptors in {\em Manduca sexta} \cite{pratt2017}.  The values are taken to be $a=\SI{5}{\milli\second}$, $b=\SI{4}{\milli\second}$, $\omega_\text{STA}=\SI{1000}{\radian\per\second}$, $c=10$, $d=0.5$, $N=\SI{40}{\milli\second}$ and $C_\xi=0.1174$.

In order to account for variations in observability due to sensor location, the surface of the wing was discretized into a 51 $\times$ 21-station grid, and the empirical observability Gramian was computed for each sensor location on the grid. Sensor measurements were taken to encode the bending strain, $\epsilon_{yy}$, and the shear strain, $\epsilon_{xy}$. The perturbation value for the initial conditions was selected as $\varepsilon=0.001$.

\begin{figure}[hbt!] 
\centering
\includegraphics[width=0.973\linewidth]{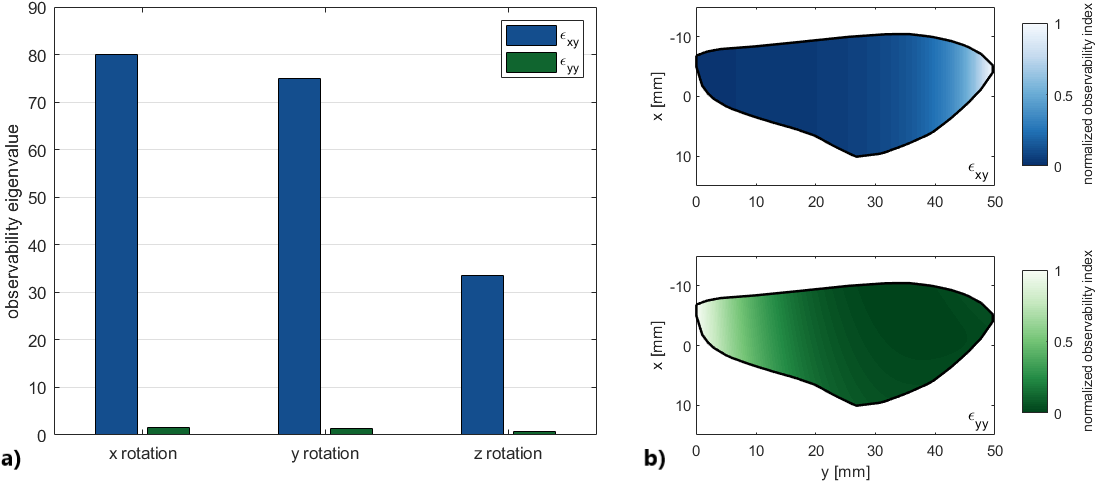}
\caption{a) Spatially averaged empirical observability
Gramian eigenvalues for rotation about the three body axes using either neural-encoded shear strain, $\epsilon_{xy}$, or bending strain, $\epsilon_{yy}$. b) Observability indices throughout the wing for shear (\textmd{\emph{top}}) and bending (\textmd{\emph{bottom}}) strain encoding, normalized individually for each sensor type.}\label{fig:3cmaps}
\end{figure}

The spatially averaged observability eigenvalues for  independent rotations about body-fixed axes are given in Fig. \ref{fig:3cmaps}a. For both shear and bending sensing, the most observable mode was the rotation about the $x$ axis while the rotation about the $z$ axis was the least observable  mode. Although the shear strain measurements resulted in relatively large eigenvalues of the empirical observability Gramian, the eigenvalues from bending strain measurements were more balanced with an average condition number of $1.9$ for bending while it was  $3.0$ for shear. The empirical observability Gramian was also utilized to study the observability of the hawkmoth using just the raw strain information, and the condition numbers of the matrix were reported in \cite{hinson2015} to be $1.2$ for bending and $1.3$ for shear. Neural encoding negatively affected the balance of eigenvalues in strain measurements but kept them within acceptable levels.

The observability index values, or simply the minimum observability eigenvalues, for bending and shear stress measurements were obtained and normalized separately. The results given in Fig. \ref{fig:3cmaps}b are consistent with the delay-free case as reported in \cite{hinson2015}. In other words, the dynamics with shear strain measurements were more observable at the wing tip while the observability index was the highest at the wing root for span-wise bending.

\subsection{Optimal Sensor Placement}
A key point about the observability of the body rotation rates is that the results are not uniform across the wing of the hawkmoth; the location of the sensors has a large effect on the observability result. Two fundamental questions can then be addressed relative to NLA sensing:
(1) optimal sensor placement from an engineering perspective to determine system design, and (2) assessment of sensor functionality in biological systems. To address the first point, a sensor placement problem can be designed to obtain an optimal configuration in terms of the sensor type (shear or bending via neural encoding) and location. With respect to the second question, possible sensor locations can be restricted to being on one of the wing veins in the hawkmoth to correspond to viable physical locations on the actual animal \cite{dickerson2014}. In either case,
a set of possible sensor locations on the veins was formed by spacing them $\SI{2}{\milli\meter}$ apart along each vein, and the empirical Gramian was computed at those locations using the same parameters in the observability analysis above.

\begin{figure}[hbt!] 
\centering
\includegraphics[scale=0.639]{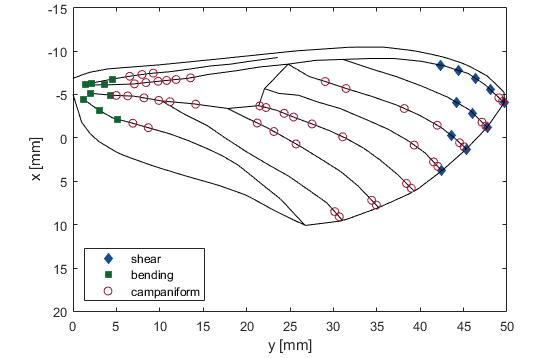}
\caption{ Optimal sensor set compared to locations of campaniform sensilla found on a hawkmoth forewing in \cite{dickerson2014}. The clusters of campaniform sensilla located at the wing root near the wing hinge are not shown.}\label{optsenplace}
\end{figure}

To set up the sensor selection framework, note that the sum of the Gramians corresponding to each sensor in a set being used individually gives the observability Gramian for all sensors in the set being used simultaneously, i.e.,
\begin{equation}
 W_o^\varepsilon(\beta) = \sum\limits_{i=1}^p \beta_i( W_o^\varepsilon)_i,
\end{equation}
where $\beta_i\in\{0,1\}$ is a sensor activation function indicating whether a given sensor is in use, and $p$ is the number of sensors in the set. By relaxing the sensor activation constraint to a linear inequality constraint, an observability-based optimal sensor placement problem can be posed as
\begin{equation}
\begin{aligned}
& \min\limits_{\beta} 
& & \kappa(W_o^\varepsilon(\beta)) + w_\nu \nu(W_o^\varepsilon(\beta)) \\
& \text{subject to}
& &0 \leq \beta \leq 1 \\
& & &\sum\beta = r, 
\end{aligned}
\end{equation}
where $r\leq p$ is the desired number of sensors to place on the wing, and $w_\nu\geq0$ is the weight. To solve this convex optimization problem, we used CVX, a package for specifying and solving convex programs \cite{cvx,gb08}. The resultant sensor location and types for $r=20$ and $w_\nu=20$ are given in Fig. \ref{optsenplace} along with physical campaniform sensilla locations.

Two clusters of sensors are obtained from the numerical optimization while there are three clusters of campaniform sensilla on the actual insects. The sensors are gathered where the $\epsilon_{xy}$ and $\epsilon_{yy}$-based observability indices are maximum. The results here have a similar pattern to the sensor locations found for the raw measurements in \cite{hinson2015} which is to be expected as their observability index behaviors across the wing are parallel.

\subsection{Varying NLA Functions}
In order to further explore the effects of neural-inspired sensing with delay, we tested different NLA functions from \cite{mohren2018}. To be consistent with that work, we only studied the observability via neural encoding of the bending strain, $\epsilon_{yy}$. This application was based on the fact that there is no evidence that campaniform sensilla are able to detect shear strain.

\begin{figure}[hbt!] 
\centering
\includegraphics[width=1\linewidth]{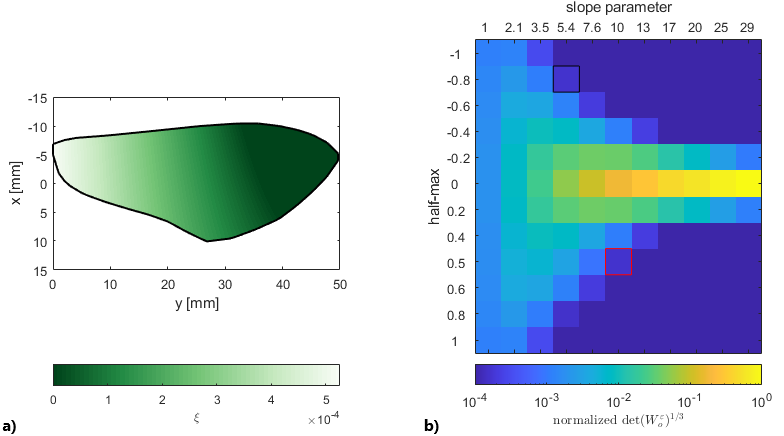}
\caption{a) The time-averaged projected stimulus ($\xi$) throughout the wing. b) The change of the spatiotemporal average value of $[\determ{W_o^\varepsilon}]^{1/3}$ by the two NLA parameters. The black and red boxes correspond to the functions given in Fig. \ref{hyptan}b. The parameters corresponding to the red box are also the ones derived experimentally.}\label{varyNLA}
\end{figure}

The resultant distribution of the time-averaged projected stimulus, $\xi$, on the wing is given in Fig. \ref{varyNLA}a. The spatiotemporal mean value is $1.61\times10^{-4}$. The values of the tested NLA parameters, $c$ and $d$, ranged from $1$ to $29$ and $-1$ to $1$, respectively, while the STA parameters duplicated the ones above. Figure \ref{varyNLA}b illustrates spatially-averaged value of the $n$-th root of the determinant for varying slope parameter and half-max values. The rotation rates were most observable when $d=0$, and the observability decreased as the half-max moved away from zero. The slope parameter, $c$, also affects the observability: As $c$ increases, the average determinant value decreases except the case when $d=0$.

Theorem \ref{theo:mainDet} and the relation of the linear observability matrix and the linear observability Gramian led us to investigate the relation between $[\determ{W_o^\varepsilon}]^{1/n}$ and $(\operatorname{d} \operatorname{NLA}(\xi)/{\operatorname{d} \xi})^2$. We observed that they are highly correlated ($\rho=0.9978$) for the 132 NLA parameter combinations, which supports the anticipated relation of output sensitivity and the derivative of composing function, $g$.

\section{Conclusion}\label{sec:last}

In this paper, we presented observability analysis approaches for two kinds of systems: systems with output delay and systems with an output function in the form of function composition. Although we were inspired by the neural encoding mechanism in animal sensing, the results are also useful for existing engineered systems as in the example of the double integrator with a saturated output. We also studied optimal neural-inspired sensor placement and the effect of neural encoding parameters on system observability. We note that there are some combinations of NLA parameters providing higher observability, which might be insightful in both computational neuroscience and engineering applications.

A possible future research direction is the analysis of systems with neural-inspired sensing in the presence of process noise where having an embedded filter would be more effective. This analysis will require the use of stochastic observability. In \cite{powelArXiv}, it is shown that such stochastic methods can reveal observability in systems that are considered unobservable using traditional deterministic tools. As another possible direction, the resultant shear strain distribution from the Euler-Lagrange model does not reflect the kinematics of the hawkmoth wing completely, and we are considering replacing the model with a finite element model, which would increase the need for numerical tools like the empirical Gramian. Finally, we are interested in analytically studying the empirical observability Gramian for the systems with a composite output function.

\section*{Appendix: Proof of Lemma 1}

We prove by induction. Our base case is $k=1$. For $k=1$, we obtain $L_{\bf f}(g\circ h)=(g^{(1)}\circ h)L_{\bf f}(h)$ from Eq.\,(\ref{eq:lemma}), which one can confirm via the chain rule for derivatives. For the induction step, we show that if Eq.\,(\ref{eq:lemma}) holds for $k$, it also holds for $k+1$. We have
\begin{align}
     L^{k+1}_{\bf f}(g\circ h) = {L_{\bf f}}L^{k}_{\bf f}(g\circ h)
      = \sum_{j=1}^k\left[  {\nabla}
     (g^{(j)}\circ h) \sum_{s\in M_{k,j}}\left(\prod_{s_i\in s} L_{\bf f}^{s_i}(h)\right) 
      {\cdot f}+(g^{(j)}\circ h)\sum_{s\in M_{k,j}}{\nabla}\left(\prod_{s_i\in s} L_{\bf f}^{s_i}(h)\right){\cdot f}\right]\,,
    \label{eq:proof_lemma2}
\end{align}
where we have used Eq.\,(\ref{eq:lemma}) and the product rule for derivatives. Applying the chain rule for derivatives to the first part of the right-hand side of Eq.\,\ref{eq:proof_lemma2} and the product rule to the gradient of $\prod L_{\bf f}^{s_i}(h)$, we obtain
\begin{align}
    L^{k+1}_{\bf f}(g\circ h)
    & = \sum_{j=1}^k\left[ (g^{(j+1)}\circ h)\, {\nabla h} \sum_{s\in M_{k,j}}\left(\prod_{s_i\in s} L_{\bf f}^{s_i}(h)\right){\cdot \bf f}+(g^{(j)}\circ h)\sum_{s\in M_{k,j}}{\sum_{s_\ell\in s}}{\nabla\left(L_{\bf f}^{s_\ell}(h)\right)}\left(\prod_{\substack{s_i\in s\,\\{i\neq\ell}}} L_{\bf f}^{s_i}(h)\right){\cdot \bf f}\right]\,,\label{eq:proof_lemma3}
\end{align}
The inner products of the gradient operators in Eq.\,\ref{eq:proof_lemma3} with $\vect{f}$ can be expressed as Lie derivatives, that is,
\begin{align}
    L^{k+1}_{\bf f}(g\circ h) = \sum_{j=1}^k\left[ (g^{(j+1)}\circ h)\, \sum_{s\in M_{k,j}}\left(\prod_{s_i\in s} L_{\bf f}^{s_i}(h)\right){L_{\bf f}(h)} +(g^{(j)}\circ h)\sum_{s\in M_{k,j}}{\sum_{s_\ell\in s}}{L_{\bf f}^{s_\ell+1}(h)}\left(\prod_{\substack{s_i\in s\,\\{i\neq\ell}}} L_{\bf f}^{s_i}(h)\right)\right]\,.
    \label{eq:proof_lemma45}
\end{align}
We shift the index $j$ in the first part of the right-hand side of Eq.\,(\ref{eq:proof_lemma45}), so that
\begin{equation}
\begin{aligned}
    L^{k+1}_{\bf f}(g\circ h)
    & = \sum_{j=2}^{k+1}\left[ (g^{(j)}\circ h)\, \sum_{s\in M_{k,j-1}}\left(\prod_{s_i\in s} L_{\bf f}^{s_i}(h)\right)L_{\bf f}(h)\right]+\sum_{j=1}^k\left[(g^{(j)}\circ h)\sum_{s\in M_{k,j}}{\color{black}\sum_{s_\ell\in s}}L_{\bf f}^{s_\ell+1}(h)\left(\prod_{\substack{s_i\in s\,\\{\color{black}i\neq\ell}}} L_{\bf f}^{s_i}(h)\right)\right]\,\\
    & = \sum_{j=1}^{k+1}\left[ (g^{(j)}\circ h)\, \sum_{s\in A\cup B}\left(\prod_{s_i\in s} L_{\bf f}^{s_i}(h)\right)\right]{ - (g^{(1)}\circ h) \sum_{M_{k,0}}\left(\prod_{s_i\in s} L_{\bf f}^{s_i}(h)\right)}{ - (g^{(k+1)}\circ h)\sum_{M_{k,k+1}}\left(\prod_{s_i\in s} L_{\bf f}^{s_i}(h)\right)}\,,
    \label{eq:proof_lemma7}
\end{aligned}
\end{equation}
where $A = \left\{s\cup\{1\}\:|\:s \in M_{k,j-1}\right\}$ and $B = \left\{s\backslash\{s_i\}\cup\{s_i+1\}\:|\:s_i \in s,\:s\in M_{k,j}\right\}$. By definition of $M_{k,j}$ (see Eq.\,(\ref{eq:mdef1})), $M_{k,0}$ and $M_{k,k+1}$ are empty sets for any $k$. Since $A\cup B$ is equal to $M_{k+1,j}$, we obtain
\begin{align}
    L^{k+1}_{\bf f}(g\circ h)
    & = \sum_{j=1}^{k+1}\left[ (g^{(j)}\circ h)\, \sum_{s\in M_{k+1,j}}\left(\prod_{s_i\in s} L_{\bf f}^{s_i}(h)\right)\right]\,.\qed \nonumber
\end{align}

\section*{Funding Sources}

This work was funded in part by the Air Force Office of Scientific Research MURI FA9550-19-1-0386 and in part by the Washington State Joint Center for Aerospace Technology Innovation.

\bibliographystyle{unsrtnat}
\bibliography{biblio}

\end{document}